\documentclass{raa}            

\usepackage{graphicx,times}             

\begin{document}

   \title{Non-detection of pulsed radio emission from magnetar Swift J1834.9$-$0846: constraint on
   the fundamental plane of magnetar radio emission
}

   \volnopage{Vol.0 (200x) No.0, 000--000}      
   \setcounter{page}{1}          

   \author{
 H. Tong, J. P. Yuan and Z. Y. Liu
\inst{}
   }

   \institute{Xinjiang Astronomical Observatory, Chinese Academy of Sciences, Urumqi, Xinjiang 830011,
   China {\it tonghao@xao.ac.cn; yuanjp@xao.ac.cn}\\
           }

   \date{Received~~2012 month day; accepted~~2012~~month day}

\abstract{
The magnetar Swift J1834.9$-$0846 is observed using Nanshan 25 meter radio telescope.
No pulsed radio emission is detected. The upper limit on pulsed radio emission from
this source is $0.5 \rm \,mJy$. According to the ``fundamental plane'' for radio magnetars,
this source should have radio emission. Therefore, our results put constraints
on the existence of a fundamental plane of magnetar radio emission.
We argue that a magnetar's ability to emit radio emission may have little to do with the spin down
luminosity and it is related to the magnetar X-­ray luminosity. The only necessary condition is a
relatively low X-­ray luminosity.
}

   \authorrunning{Tong, Yuan \& Liu}            
   \titlerunning{Non-detection of pulsed radio emission from magnetar Swift J1834.9$-$0846}  

   \maketitle

%
%
\section{Introduction}           

Magnetars are assumed to be neutron stars powered by strong magnetic field decay (Duncan \& Thompson 1992).
They form a different pulsar population from that of rotation-powered pulsars. 
Normal rotation-powered pulsars are usually radio emitters. 
Radio emitting rotation-powered pulsars are commonly known as radio pulsars.
However, magnetars manifest themselves mainly as anomalous X-ray pulsars (AXPs) and 
soft gamma-ray repeaters (SGRs)(Tong et al. 2010, 2011).
Until recently, no radio pulsations had been
observed from any of the magnetars. The discovery of transient pulsed radio emission from one magnetar
has bridged the gap between radio pulsars and magnetars (Camilo et al. 2006). Up to now,
more than twenty magnetars have been discovered \footnote{McGill online catalog:
http://www.physics.mcgill.ca/$\sim$pulsar/magnetar/main.html}. Three of them are radio-loud magnetars
(Camilo et al. 2006, 2007; Levin et al. 2010).

Recently, Rea et al. (2012) tried to understand magnetar radio emission from an empirical point of view.
They proposed that magnetars are radio-loud if and only if their quiescent X-ray luminosities are smaller
than their rotational energy loss rates: $L_{\rm qui} < \dot{E}$. This is the key point of the
``fundamental plane'' of magnetar radio emission.
Since Rea et al. (2012) published their paper, there are two new sources up to now: SGR Swift J1822.3$-$1606 and
SGR Swift J1834.9$-$0846. For the young magnetar Swift J1834.9$-$0846, the upper limit of its quiescent
X-ray luminosity is lower than its rotational energy loss rate (Kargaltsev et al. 2012). This source
should be another radio-loud magnetar
if the fundamental plane of magnetar radio emission proposed by Rea et al. (2012) is correct.

\subsection{X-ray Observations of Swift J1834.9$-$0846}

According to Kargaltsev et al. (2012), Swift J1834.9$-$0846 has a rotation period of $2.48 \,\rm s$
and a period derivative $\dot{P}=0.796\times 10^{-11}$ { s~s$^{-1}$}. Its characteristic magnetic field is
$B=3.2\times 10^{19} \sqrt{P\dot{P}} = 1.4\times 10^{14} \,\rm G$. It may also be associated with the
supernova remnant W41 (Tian et al. 2007; Kargaltsev et al. 2012).
Therefore, Swift J1834.9$-$0846 is similar to the radio
emitting magnetar AXP 1E 1547.0--5408 (Camilo et al. 2007): they have similar rotation period and similar characteristic
magnetic field, and both are young sources in association with supernova remnants. The rotational energy
loss rate of Swift J1834.9$-$0846 is $\dot{E}= 3.95 \times 10^{46} \dot{P} P^{-3} =2.1\times 10^{34} \,\rm erg \, s^{-1}$.
According to figure 3 in Kargaltsev et al. (2012), Swift/XTR observed a declining flux of Swift J1834.9$-$0846.
From figure 3 there, the upper limit of the source's quiescent flux is:
$f_{\rm qui} < 3\times 10^{-12} \,\rm erg \, cm^{-2}\, s^{-1}$. The corresponding upper limit of the source's quiescent
luminosity is $L_{\rm qui} = 4\pi d^2 f_{\rm qui}<5.7\times 10^{33} d_{4}^2 \, \rm erg \, s^{-1}$.
Here the source distance is chosen as $4\,\rm kpc$,
considering its potential association with the supernova remnant W41 (Tian et al. 2007; Kargaltsev et al. 2012).
Furthermore, pre-outburst XMM and Chandra  observations showed that the source flux
was about 100 and 1000 times smaller, respectively
(Section 3.3.2 in Younes et al. 2012; Section 6.2.3 in Kargaltsev et al. 2012).
Therefore, the quiescent luminosity of Swift J1834.9$-$0846 must be smaller than its rotational energy loss rate.

This source meets all the criteria of the fundamental plane of magnetar radio emission from Rea et al. (2012).
\begin{enumerate}
 \item Quiescent X-ray luminosity smaller than the rotational energy loss rate.
 \item A high acceleration potential along pulsar open field line regions. For Swift J1834.9$-$0846,
 the corresponding acceleration potential is:
 $\Delta V = 4.2\times 10^{20} \sqrt{\dot{P}/P^3} = 3.0\times 10^{14} \, \rm Volts$.
 \item Burst/outburst to trigger the radio emission. Swift J1834.9-0846 showed SGR-type burst on 2011 August 7
 (D'Elia et al. 2011).
 Its declining flux points to a likely recent outburst.
 \item It lies relatively nearby. A possible distance of $4 \,\rm kpc$ is obtained considering its potential association
 with supernova remnant W41.
\end{enumerate}
If the fundamental plane proposed by Rea et al. (2012) was correct, Swift J1834.9$-$0846
should have radio emission. And from previous experience of magnetar radio emission, we should detect
its radio emission in recent years. Therefore, Swift J1834.9$-$0846 provides us the first
opportunity to test the ``fundamental plane'' of magnetar radio emission.

Radio observations, data analysis and results are presented in section 2.
Discussions and conclusions are given in section 3.

\section{Radio observations, data analysis and results}

We have observed this source for $2\times1$ hours, with one hour folding mode and the other searching mode. 
Here, we mainly discuss the searching observation and data analysis.
The searching data were first checked for the presence of non-dispersed radio frequency interference (RFI).
Interference signals above our set 5$\sigma$ threshold level were removed from the raw data prior to our analysis.

\subsection{Radio Observations of Swift J1834.9$-$0846}

The observation of Swift J1834.9$-$0846 was made using Nanshan 25 meter radio telescope
(Xinjiang Astronomical Observatory)
on 24 June, 2012 for 1.0 hour. Data were taken with the cryogenic receiver at a center frequency of 1540 MHz.
A bandwidth of 320 MHz was used in our observation, split into 128 continuous 2.5 MHz frequency channels (Wang et al. 2001).
Dual linear polarizations were summed, and the frequency channels were one-bit sampled every 1.0 ms.
The data were recorded on computer disk and transferred from the observatory to a linux computing server for processing.
No pulsed radio emission is detected \footnote{We observed this source for 1.0 hour in May 2012 in folding mode,
which is also not detected.}.

\subsection{Folding Search}

The data were analysed using the  pulsar signal processing package {\sc SIGPROC}\footnote{See http://sigproc.sourceforge.net} (Lorimer et al. 2000). 
The maximum dispersion measure (DM) detected in the known pulsars is 1456 pc cm$^{-3}$.
According to the NE2001 model for the galactic distribution of free electrons (Cordes \& Lazio 2002),
Swift 1834.9$-$0846 has a DM of 197 pc cm$^{-3}$ assuming the distance of 4 kpc.
So the data were de-dispersed using 750 trial DMs ranging from 0 to 1500 pc cm$^{-3}$. For each DM trial, the full 320 MHz of bandwidth was de-dispersed.

A barycentric folding period was determined using measurement from previous X-ray observation of Swift J1834.9$-$0846.
With the reported period and its associated uncertainties from the ${\it RXTE}$ observations,
we extrapolated the period to the 24 June, 2012 epoch and determined the barycentric folding period to be 2.48249785(10) s.

Periods $\pm$5 ms ($\sim$0.2\%) from the nominal period were searched with a step of 0.01 ms.
Each of these folding trials was conducted for DMs between 0 and 1500 pc cm$^{-3}$ with steps 2 pc cm$^{-3}$.
A total of 375,000 DM and period combinations were tried (750 DM trials, and 500 folds per DM trial),
and for each trial the $\chi ^2$ significance of the folded profile was recorded. We chose 5$\sigma$ as a reasonable threshold for the signal,
and we found no convincing pulsar candidates at the 5$\sigma$ significance level or higher in the search. 
At the same time, we also searched for periodic signal using FFT with SEEK program at each tried DM. 
No pulsar candidate was found in this observation.

\subsection{Upper Limit on Pulsed Radio Emission}

The minimum detectable flux density can be given by the function (Manchester et al. 1996)
\begin{equation}
S_{\rm min}=\frac {\alpha \beta T_{\rm sys}}{G \sqrt{N_{\rm p}\Delta t\Delta\nu}}\sqrt{\frac{W_{\rm e}}{P-W_{\rm e}}}.
\end{equation}
It is affected by system noise $T_{\rm sys}$, telescope gain $G$, 
number of polarizations $N_{\rm p}$, integration time $\Delta t$, receiver bandwidth $\Delta \nu$, 
assumed signal to noise ratio $\alpha$, digitization and other processing losses $\beta$, 
pulsar period $P$ and effective pulse width $W_{\rm e}$. 
The effective pulse width depends on the intrinsic pulse width $W$, the sampling time $\delta t$, 
the dispersion smearing time across one sub-channel $\delta t_ {\rm DM}$ and 
the interstellar scattering boarding $\delta t_{\rm scatt}$
\begin{equation}
W_{\rm e}=\sqrt{W^2+\delta t^2+\delta t^2_{\rm DM}+\delta t^2_{\rm scatt}},
\end{equation}
where $\delta t =1 \,\rm ms$, $\delta t_{\rm DM} = 8.3 \times 10^6 \, DM \, \nu^{-3}_{\rm MHz} \, B \, \rm ms = 1.136 \,\rm ms$, 
$\delta t_{\rm scatt} = (\frac{DM}{1000})^{3.5}  (\frac{400} {\nu_{\rm MHz}})^4 \,10^3\,\rm ms = 0.016\,\rm ms$ 
with central frequency $\nu = 1540 \,\rm MHz$, 
$DM = 200\, \rm {cm}^{-3} \, pc$, and bandwidth $B = 2.5 \,\rm MHz$ (Lyne \& Graham-Smith 2012). 
The intrinsic pulse width was assumed to be $0.05\,P$ (Lazarus et al. 2012). 
The effective pulse width will be $0.05\,P$ when the sum of the three contributions 
is much less than the intrinsic pulse width. 
For Nanshan 25 meter radio telescope pulsar observing system ($T_{\rm sys}$ = 40 K, $G$ = 0.1 K Jy$^{-1}$, 
$N_{\rm p}$ = 2, $\Delta t$ = 3600 s, $\Delta \nu= 320 $ MHz, $\beta$ = 1.5), 
we set $\alpha$ equal 5. 
Then, we obtain an upper limit on pulsed radio emission from Swift 1834.9$-$0846 as 0.5 mJy.

\section{Discussions and conclusions}

According to Rea et al. (2012), a magnear will emit radio pulsations
if and only if its quiescent X-ray luminosity is smaller than its rotational energy loss rate: $L_{\rm qui} < \dot{E}$.
Swift J1834.9$-$0846 meets all the criteria of the fundamental plane of magnetar radio emission.
If the proposal of Rea et al. (2012) was correct, then it should have radio emission.
However, we detect no pulsed radio emission from this source.
It may be that magnetars with $L_{\rm qui} < \dot{E}$ might have radio emission. At the same time,
they can also have no radio emission. Despite the original proposal of Rea et al. (2012),
$L_{\rm qui} < \dot{E}$ may only be a necessary condition for a magnetar to emit radio pulsations.
In the Rea et al. (2012) paper, two of five sources with
$L_{\rm qui} < \dot{E}$ have no radio emission. With the addition of Swift J1834.9$-$0846, a total of
six sources have $L_{\rm qui} < \dot{E}$. Three of them have radio emissions (AXP XTE J1810$-$4197,
AXP 1E 1547.0$-$5408, and PSR J1622$-$4950). While the other three are not detected in radio (PSR J1846$-$0258,
SGR 1627$-$41, and Swift J1834.9$-$0846). It is possible that the three later magnetars are actually
radio emitting sources. Their radio emissions could have been missed because of beaming, absorption due to the
environment or large distances etc (Rea et al. 2012).
However, we also want to highlight the possibility that they do not have any radio emission at all.

The X-ray emissions of magnetars are magnetism-powered.
They have nothing to do with the rotational energy loss rate (Thompson et al. 1996, 2002).
However, the fundamental plane of magnetar radio emission links the magnetar quiescent X-ray luminosity
with its rotational energy loss rate. Rea et al. (2012) did this since they believed that the radio emissions
of magnetars are rotation-powered. However, the characteristics of magnetar radio emission are very different
from that of radio pulsars (Mereghetti 2008): a variable flux and pulse profile, a flat spectra, and most
importantly it is transient in nature. If the radio emissions of magnetars are rotation-powered, the same as
that of radio pulsars, we should see similar radio emission properties in radio magnetars and radio pulsars.
However, this is not what has been observed (Camilo et al. 2006, 2007; Levin et al. 2010).
We find it reasonable to think that the magnetar radio emission
comes from a different energy reservoir. In the case of magnetars, the natural energy budget
is the magnetic energy. Therefore, the radio emissions of magnetars may be magnetism-powered
instead of rotation-powered. The X-ray emissions of magnetars can vary significantly. Then it is not surprising
that their radio emissions are also variable, since they are from the same energy reservoir.

For some magnetars, they can also have relative short period
(e.g., AXP 1E 1547.0$-$5408 has a period of $2.1\,\rm sec$, Camilo et al. 2007).
The rotational energy is also very significant. Therefore, we may also expect some
rotation-powered activities in magnetars (Zhang 2003). For example, there may also exist rotation-powered
radio emissions in magnetars. Therefore, there could be two types of radio emissions in magnetars:
magnetism-powered radio emission and rotation-powered radio emission. At present, only transient pulsed
radio emissions are observed in magnetars. They are more likely to be magnetism-powered. In the future,
more radio-loud magnetars will be discovered (e.g., by FAST or SKA). Among them, we may also see
pulsed non-transient radio emissions in some magnetars, with properties similar to that of ordinary radio pulsars.
These radio emissions may be rotation-powered.

At present, three of six sources with $L_{\rm qui} < \dot{E}$ have radio emissions. One physical
reason is that: low luminosity magnetars tend to have similar magnetospheric structure as that
of radio pulsars (Section 4.2 in Tong et al. 2012). The coherent radio emission condition is
more likely to be fulfilled in low luminosity magnetars. Only a relatively low X-ray luminosity
is required. It has few relation with the magnitude of rotational energy loss rate.
Therefore, the ``fundamental plane'' of magnetar radio emission
(if it exists\footnote{See also Ho (2012) for criticisms on the existence of fundamental plane of magnetar radio emission.})
should be: ``low luminosity magnetars are more likely to have radio emissions''.
Nothing further can be said.
Since the Nanshan telescope is relatively small (25 meters in diameter),
only a crude upper limit is obtained.
Continued monitoring at other radio telescopes is highly recommended\footnote{During
the submission process of this paper, Esposito et al. (arXiv:1212.1079) reported their GBT
radio observations of Swift J1834.9$-$0846 which is also not detected. Their observations
were made from August 2011 to November 2011.
This result is consistent with our analysis here. However, a different explanation is discussed there.}.

In summary, previously Rea et al. (2012) said that the necessary and sufficient condition for a magnetar 
to emit in radio is $L_{\rm qui} < \dot{E}$ (quiescent X-ray luminosity smaller than the 
rotational energy loss rate), and the radio emission is powered by the star's rotational energy.  
Based on our observation on Swift J1834.9$-$0846, we want to point out an alternative possibility. 
\begin{enumerate}
 \item Since Swift J1834.9$-$0846 is not detected in radio (in contradiction with the prediction of Rea et al. 2012),
 $L_{\rm qui} < \dot{E}$ may only be a necessary condition. 
 \item Considering the differences between radio emissions of magnetars and normal radio pulsars, it is 
 more reasonable to think that radio emissions of known magnetars are powered by their magnetic energy 
 (instead of rotational energy proposed by Rea et al. 2012). 
 \item If the radio emissions of magnetars are powered by their magnetic energy, and combining theoretical studies 
 of magnetar magnetosphere (Tong et al. 2012), then the necessary condition for a magnetar to have radio emission
 may have little to do with its rotational energy loss rate. Only a relatively low X-ray luminosity is required. 
 \end{enumerate}

\section*{Acknowledgments}
The authors would like to thank the Referee for helpful suggestions and N. Rea for comments.
H. Tong would like to thank KIAA at PKU for support of visiting. 
This work is supported by Natural Science Foundation of Xinjiang (NO.2009211B35), 
National Natural Science Foundation of China (11103021, 11173041, 10903019),
West Light Foundation of CAS (XBBS201021, LHXZ201201).

\label{lastpage}

\end{document}